\documentclass[conference]{IEEEtran}
\usepackage{cite}
\usepackage{amsmath,amssymb,amsfonts}
\usepackage{algorithmic}
\usepackage{graphicx}
\usepackage{amssymb,comment}
\usepackage{algorithmic}
\usepackage{algorithm}
\usepackage{caption} 
\usepackage{subfig}
\usepackage{subcaption}
\usepackage{scrextend}
\usepackage{verbatim}
\usepackage{color}
\usepackage{tablefootnote}
\begin{document}
	
	\newtheorem{theorem}{\bf Theorem}
	\newtheorem{lemma}{\bf Lemma}
	\newtheorem{remark}{\bf Remark}
	\newtheorem{proposition}{\bf Proposition}
	\newcommand{\bs}[1]{\boldsymbol{#1}}
	\newcommand{\mbf}[1]{\mathbf{#1}}
	\newcommand{\ib}[1]{\in\mathbb{#1}}
	\newcommand{\ic}[1]{\in\mathcal{#1}}
	\newcommand{\ca}[1]{\mathcal{#1}}
	
	\title{Movable Antennas Meet Intelligent Reflecting Surface: When Do We Need Movable Antennas?}
	\author{\IEEEauthorblockN{Xin Wei\IEEEauthorrefmark{1}, Weidong Mei\IEEEauthorrefmark{1}, Qingqing Wu\IEEEauthorrefmark{2}, Boyu Ning\IEEEauthorrefmark{1}, and Zhi Chen\IEEEauthorrefmark{1}}
		\IEEEauthorblockA{\IEEEauthorrefmark{1}National Key Laboratory of Wireless Communications,\\
		University of Electronic Science and Technology of China, Chengdu, China.}
		\IEEEauthorblockA{\IEEEauthorrefmark{2}Department of Electronic Engineering, Shanghai Jiao Tong University, Shanghai, China.}
		Email: xinwei@std.uestc.edu.cn, wmei@uestc.edu.cn, qingqingwu@sjtu.edu.cn,\\ boydning@outlook.com, chenzhi@uestc.edu.cn}
	
	\maketitle
	
	\begin{abstract}
		Intelligent reflecting surface (IRS) and movable antenna (MA)/fluid antenna (FA) techniques have both received increasing attention in the realm of wireless communications due to their ability to reconfigure and improve wireless channel conditions. In this paper, we investigate the integration of MAs/FAs into an IRS-assisted wireless communication system. In particular, we consider the downlink transmission from a multi-MA base station (BS) to a single-antenna user with the aid of an IRS, aiming to maximize the user's received signal-to-noise ratio (SNR), by jointly optimizing the BS/IRS active/passive beamforming and the MAs' positions. Due to the similar capability of MAs and IRS for channel reconfiguration, we first conduct theoretical analyses of the performance gain of MAs over conventional fixed-position antennas (FPAs) under the line-of-sight (LoS) BS-IRS channel and derive the conditions under which the performance gain becomes more or less significant. Next, to solve the received SNR maximization problem, we propose an alternating optimization (AO) algorithm that decomposes it into two subproblems and solve them alternately. Numerical results are provided to validate our analytical results and evaluate the performance gains of MAs over FPAs under different setups.
	\end{abstract}
	
	\section{Introduction}
	The intelligent reflecting surface (IRS) is regarded as a promising technique for future sixth-generation (6G) wireless networks due to its superior characteristics such as channel reconfiguration capability, low-power consumption, and low-cost deployment \cite{Q_Wu_PIEEE}. By individually adjusting the reflecting coefficient of each reflecting element, the IRS can jointly alter the strength/direction of its reflected signal for achieving various objectives, such as signal enhancement, interference nulling, and spatial multiplexing among them \cite{Q_Wu_Towards,W_Mei_PIEEE}.
	
	On the other hand, the movable antenna (MA)/fluid antenna (FA) technology has drawn great interest in academia and industry recently. Similar to the IRS, the MA/FA technology can reconfigure wireless channel conditions by enabling multiple antennas to be flexibly moved within a confined region at the transmitter/receiver \cite{B_Ning_Arte,L_Zhu_COMMAG,C_Wang_WMAG}. As such, both IRS and MA/FA technologies have been extensively studied in the literature under various system setups, e.g., multiple-input single-output (MISO) \cite{Q_Wu_TWC, N_Li_NOMA,W_Mei_Graph}, multiple-input multiple-output (MIMO) \cite{S_Zhang_MIMO,W_Ma_MA_MIMO}, spectrum sharing \cite{X_Guan_WCL,X_Wei_WCL}, physical-layer security \cite{M_Cui_Secure,B_Feng_PHY}, and so on. Furthermore, some recent works have also proposed a new MA-inspired IRS architecture with movable reflecting elements and characterized its performance limits \cite{G_Hu_ME,Y_Zhang_ME}. However, to the best of our knowledge, there are no existing works focusing on the integration of MAs into an IRS-assisted wireless system. Particularly, due to the comparable capability of MAs and IRS for channel reconfiguration, it remains an open problem whether the performance gain of MAs still exists or not in the presence of the IRS's passive beamforming.
	
	\begin{figure}[!t]
		\centering
		\captionsetup{justification=raggedright,singlelinecheck=false}
		\centerline{\includegraphics[width=0.4\textwidth]{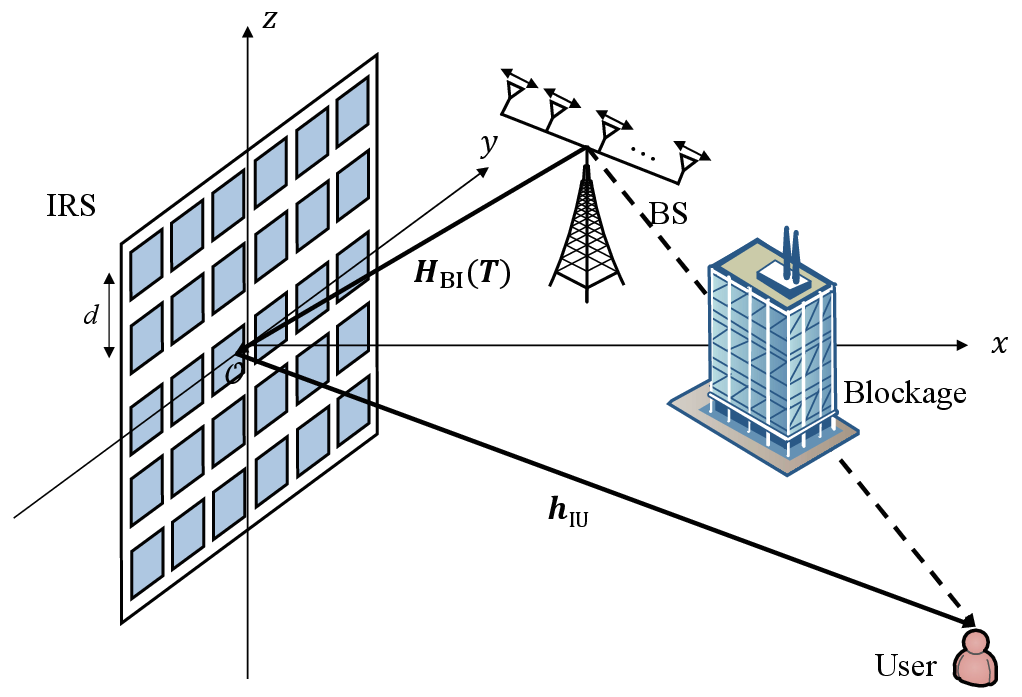}}
		\captionsetup{font=footnotesize}
		\caption{MA-enhanced IRS-aided MISO communication system.}
		\label{Fig_SysModel}
		\vspace{-12pt}
	\end{figure}
	
	To fill in this gap, we investigate in this paper an MA-enhanced IRS-assisted MISO communication system, where a base station (BS) equipped with multiple MAs communicates with a user with the aid of a fixed IRS. We aim to maximize the received signal-to-noise ratio (SNR) at the user by jointly optimizing the active/passive beamforming at the BS/IRS and the MAs' positions, which appears to be a highly non-convex problem that is challenging to be optimally solved. Due to the similar capabilities of MAs/IRS for channel reconfiguration, we first conduct theoretical analyses of the performance gain of MAs over conventional fixed-position antennas (FPAs) under the line-of-sight (LoS) BS-IRS channel to drive essential insights. It is unveiled that the performance gain of MAs over FPAs vanishes in the case of a single MA or far-field BS-IRS channel condition. Next, we propose an alternating optimization (AO) algorithm to solve the primal SNR maximization problem by decomposing it into two subproblems and solve them alternately. Numerical results are provided to validate our theoretical analyses and evaluate the performance gains of MAs over FPAs under different setups.
	
	\begingroup
	\allowdisplaybreaks
	\section{System Model and Problem Formulation}
	As shown in Fig. 1, we consider a MISO wireless communication system, where a fixed IRS is deployed to assist in the communication from the BS to the user. The direct channel from the BS to the user is assumed to be severely blocked and thus can be ignored in this paper. We assume that the BS is equipped with a linear array of $N$ ($N\ge1$) MAs, while the user is equipped with a single FPA. For convenience, we establish a global three-dimensional (3D) coordinate system, where the IRS is assumed to be located in the $yOz$-plane and centered at the origin, as shown in Fig. 1. The total number of IRS reflecting elements is denoted as $M=M_yM_z$, with $M_y$ and $M_z$ denoting the number of reflecting elements along the $y$- and $z$-axes, respectively. Without loss of generality, we assume that $M_y$ and $M_z$ are both even numbers. As such, the coordinates of the $(m_y,m_z)$-th IRS reflecting element is given by $\bs{e}_{m_y,m_z}=[0,m_yd,m_zd]^T$, where $m_y\ic{M}_y\triangleq\{0,\pm1,\pm2,\cdots,\pm\frac{M_y}{2}\}$, $m_z\ic{M}_z\triangleq\{0,\pm1,\pm2,\cdots,\pm\frac{M_z}{2}\}$. Let $\bs{\Psi}=\text{diag}\left([e^{j\varphi_1},e^{j\varphi_2},\cdots,e^{j\varphi_M}]\right)\ib{C}^{M\times M}$ denote the reflection matrix of the IRS, where $\varphi_m$ denotes the phase shift of the $m$-th reflecting element, $m\ic{M}\triangleq\{1,2,\cdots,M\}$.
	
	Furthermore, we assume that the MAs at the BS can be flexibly moved within a linear array of length $A$ meters (m), denoted as ${\cal C}_t$. The coordinate of its center is denoted as $\bs{q}_B=[x_B,y_B,z_B]^T$. Let $\bs{t}_n=[x_n,y_n,z_n]^T$ denote the coordinate of the $n$-th MA, $n\ic{N}\triangleq\{1,2,\cdots,N\}$, and $\bs{T}=[\bs{t}_1,\bs{t}_2,\cdots,\bs{t}_N]^T\ib{R}^{3\times N}$ denote the antenna position vector (APV) of the $N$ MAs.
	
	Let $\bs{H}_{\text{BI}}(\bs{T})\ib{C}^{M\times N}$ and $\bs{h}_{\text{IU}}\ib{C}^{M\times1}$ denote the BS-IRS and IRS-user channels, respectively. Note that the BS-IRS channel depends on the APV, i.e., $\bs{T}$, while $\bs{h}_{\text{IU}}$ is regardless of it. The transmit beamforming vector of the BS is denoted as $\bs{w}\ib{C}^{N\times 1}$ with $||\bs{w}||_2=1$. Then, the received signal at the user is given by 
	\begin{equation}\label{eqn_Signal}
		\begin{aligned}
			y=\sqrt{P}\bs{h}_{\text{IU}}^H\bs{\Psi}\bs{H}_{\text{BI}}(\bs{T})\bs{w}x+n,
		\end{aligned}
	\end{equation}
	where $x$ is the transmitted data symbol with $\mathbb{E}[|x|^2]=1$,  $n\sim\mathcal{CN}(0,\sigma^2)$ represents the received noise at the user with $\sigma^2$ denoting its average power, and $P$ is the BS's transmit power. Based on \eqref{eqn_Signal}, the received SNR at the user is given by
	\begin{equation}\label{eqn_SNR}
		\gamma(\bs{T},\bs{\Psi})=\frac{P}{\sigma^2}\left|\bs{h}_{\text{IU}}^H\bs{\Psi}\bs{H}_{\text{BI}}(\bs{T})\bs{w}\right|^2.
	\end{equation}
	Our objective is to maximize (\ref{eqn_SNR}) by jointly optimizing the BS's transmit beamforming $\bs{w}$, the IRS's reflection matrix $\bs{\Psi}$ and the APV $\bs{T}$. Hence, the optimization problem is formulated as
	\begin{subequations}\label{eqn_OptPrblm_P1}
		\begin{align}
			{\text{(P1)}}\quad &\underset{\bs{\Psi},\bs{T},\bs{w}}{\max}\quad \gamma(\bs{T},\bs{\Psi}) \nonumber
			\\
			\mathrm{s.t.}\quad & \varphi_m\in[0,2\pi], \forall m \ic{M}, \label{eqn_IRS_Phase_Cons} \\
			& \bs{t}_n \ic{C}_t, n \ic{N}, \label{eqn_MA_Region_Cons} \\
			& ||\bs{t}_i - \bs{t}_j|| \ge D_{\min}, \forall i,j\ic{N}, i\ne j, \label{eqn_MA_Coordinate_Cons}\\
			& ||\bs{w}||=1,
		\end{align}
	\end{subequations}
	where $D_{\min}$ denotes the minimum spacing between any two MAs to avoid mutual antenna coupling. Here, we assume that all required channel state information is available by applying the existing channel estimation techniques dedicated to MAs/IRSs, e.g., \cite{W_Ma_Compressed} and \cite{B_Zheng_Survey}.
	
	However, (P1) is a non-convex optimization problem that is challenging to solve due to the spacing constraints for MAs (i.e., \eqref{eqn_MA_Coordinate_Cons}) and the unit-modulus constraints for IRS passive beamforming (i.e., \eqref{eqn_IRS_Phase_Cons}). To reveal essential insights, we first conduct performance analyses in the next section to evaluate the fundamental gains of integrating MAs into an IRS-aided wireless communication system.
	
	\section{Performance Analysis: MAs versus FPAs}
	To facilitate our performance analyses, we first consider that the BS is equipped with a single MA, i.e., $N=1$, and characterize the maximum BS-user end-to-end channel power gain over different positions within the transmit region $\ca{C}_t$ under the optimal IRS passive beamforming. As a result, the APV reduces to a column vector, $\bs{t}\ib{R}^{3\times1}=[x_t, y_t, z_t]^T$. Moreover, we consider that the IRS can achieve a LoS-dominant channel with the BS, which usually holds in practice if the IRS is deployed in the vicinity of the BS \cite{Y_Huang_Empowering}. To capture the close BS-IRS distance in this case, we consider in this section a general near-field BS-IRS channel model, which is given by \cite{Y_Liu_NF_Review}
	\begin{equation}\label{eqn_RecPow}
		\bs{h}_{\text{BI}}(\bs{t})=\left[\frac{\lambda}{4\pi D(\bs{t},m_y,m_z)}e^{j\frac{2\pi}{\lambda}D(\bs{t},m_y,m_z)}\right]_{m_y\ic{M}_y,m_z\ic{M}_z},
	\end{equation}
	where $D(\bs{t},m_y,m_z)=||\bs{t}-\bs{e}_{m_y,m_z}||$ denotes the distance between the MA and the $(m_y,m_z)$-th IRS reflecting element. Thus, the effective BS-user channel power gain at the user can be expressed as $G_u(\bs{t},\bs{\Psi})=|\bs{h}_{\text{IU}}^H\bs{\Psi}\bs{h}_{\text{BI}}(\bs{t})|^2$. Note that since the near-field channel model is more general than its far-field counterpart, the analytical results derived from the near-field model should also hold for the far-field scenario. In contrast, the analytical results specific to the far-field model will be discussed in Section III-C.
	
	To maximize $G_u(\bs{t},\bs{\Psi})$ for any given $\bs{t}$, the optimal IRS reflection matrix should guarantee that the BS-IRS channel and the IRS-user channel are aligned, i.e.,
	\begin{equation}\label{eqn_OptIRS_Phase}
		\left[\bs{\Psi}\right]_{m,m}=\arg\left([\bs{h}_{\text{IU}}]_m\right)-\arg\left([\bs{h}_{\text{BI}}(\bs{t})]_m\right), \forall m \ic{M},
	\end{equation}
	where $[\bs{h}_{\text{IU}}]_m$  and $[\bs{h}_{\text{BI}}(\bs{t})]_m$ denote the $m$-th entries of $\bs{h}_{\text{IU}}$ and $\bs{h}_{\text{BI}}(\bs{t})$, respectively. With \eqref{eqn_OptIRS_Phase}, the effective BS-user channel power gain can be expressed as
	\begin{equation}\label{eqn_RecPow_OptPh}
		G_u(\bs{t})=\left(\frac{\lambda}{4\pi }\right)^2\left|\sum_{m_y=1}^{M_y}\sum_{m_z=1}^{M_z}\frac{|h_{\text{IU},m_y,m_z}|}{D(\bs{t},m_y,m_z)}\right|^2,
	\end{equation}
	where $h_{\text{IU},m_y,m_z}$ denotes the channel from the $(m_y,m_z)$-th IRS reflecting element to the user. In the following, we first analyze the optimal antenna position $\bs{t}$ that maximizes \eqref{eqn_RecPow_OptPh} and then analyze the fluctuation of \eqref{eqn_RecPow_OptPh} within $\ca{C}_t$.
	
	\subsection{Optimal Antenna Position}
	To simplify \eqref{eqn_RecPow_OptPh}, we introduce the following lemma.
	
	\begin{lemma}
		Let $R(\bs{t})=\sqrt{x_t^2+y_t^2+z_t^2}$ denote the distance between the MA and the origin. If $\frac{y_td}{R^2(\bs{t})},\frac{z_td}{R^2(\bs{t})}\ll1$,\footnote{Note that these conditions usually hold in practice, as $d$ is wavelength-level and should be much smaller than $R(\bs{t})$, especially for high-frequency wireless communication systems (e.g, millimeter wave).} the channel power gain in \eqref{eqn_RecPow_OptPh} can be approximated as
		\begin{equation}\label{eqn_RecPow_Approx}
			G_u(\bs{t})\approx\left(\frac{\lambda}{4\pi }\right)^2H^2(M_y,M_z,\bs{t}),
		\end{equation}
		where
		\begin{equation}\label{eqn_G}
			H(M_y,M_z,\bs{t})=\sum_{m_y=1}^{M_y}\sum_{m_z=1}^{M_z}\frac{|h_{\text{IU},m_y,m_z}|}{\sqrt{R^2(\bs{t})+(m_yd)^2+(m_zd)^2}}.
		\end{equation}
	\end{lemma}
	\begin{IEEEproof}
		The distance between the MA and the $(m_y,m_z)$-th element can be approximated as
		\begin{equation}\label{eqn_Dist_approx}
			\begin{aligned}
				&D(\bs{t},m_y,m_z)=||\bs{t}-\bs{e}_{m_y,m_z}||\\
				&=\sqrt{x_t^2+(y_t-m_yd)^2+(z_t-m_zd)^2}\\
				&=\sqrt{R^2(\bs{t})-2(y_tm_yd+z_tm_zd)+(m_yd)^2+(m_zd)^2}\\
				&\approx\sqrt{R^2(\bs{t})+(m_yd)^2+(m_zd)^2},
			\end{aligned}
		\end{equation}
		where the approximation is due to $\frac{y_td}{R^2(\bs{t})},\frac{z_td}{R^2(\bs{t})}\ll1$. By substituting \eqref{eqn_Dist_approx} into \eqref{eqn_RecPow_OptPh}, we can obtain \eqref{eqn_RecPow_Approx}.
	\end{IEEEproof}
	
	Based on \eqref{eqn_RecPow_Approx} and \eqref{eqn_G}, it is not difficult to see that $G_u(\bs{t})$ monotonically decreases with $R(\bs{t})$. Thus, if the single MA is deployed at
	\begin{equation}\label{eqn_NearestPos}
		\bs{t}^{\star} = \arg\underset{\bs{t}\ic{C}_t}{\min}\,\,R(\bs{t}),
	\end{equation}
	the channel power gain in \eqref{eqn_RecPow_Approx} can always be maximized. Note that the optimal antenna position in \eqref{eqn_NearestPos} is independent of the IRS-user channel, $\bs{h}_{\text{IU}}$.  For example, assuming that the MA array is parallel to the $x$-axis, it can be shown that $\bs{t}^{\star}=\left[-\frac{A}{2}+x_B,y_B,z_B\right]^T$. As the positions of the IRS and BS are generally fixed in practice, the optimal antenna position in \eqref{eqn_NearestPos} is fixed as well. This indicates that in the case of a single MA, it always yields the maximum SNR at the user by deploying an FPA at \eqref{eqn_NearestPos}, regardless of the IRS-user channel.
	
	\begin{figure}[!t]
		\centering
		\subfloat[]{
			\includegraphics[width=0.50\linewidth]{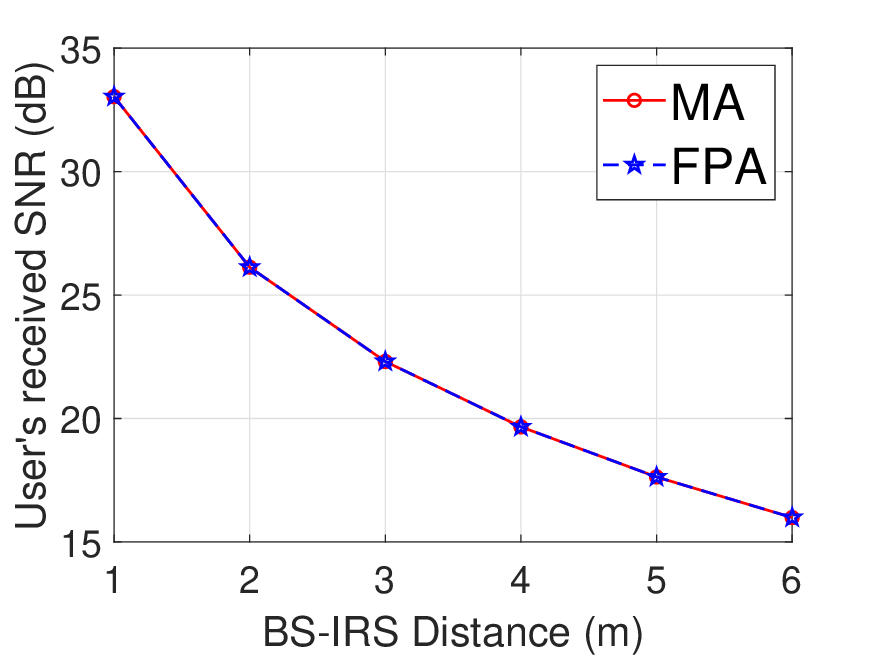}
		}
		\hfil
		\hspace{-20pt}
		\subfloat[]{
			\includegraphics[width=0.50\linewidth]{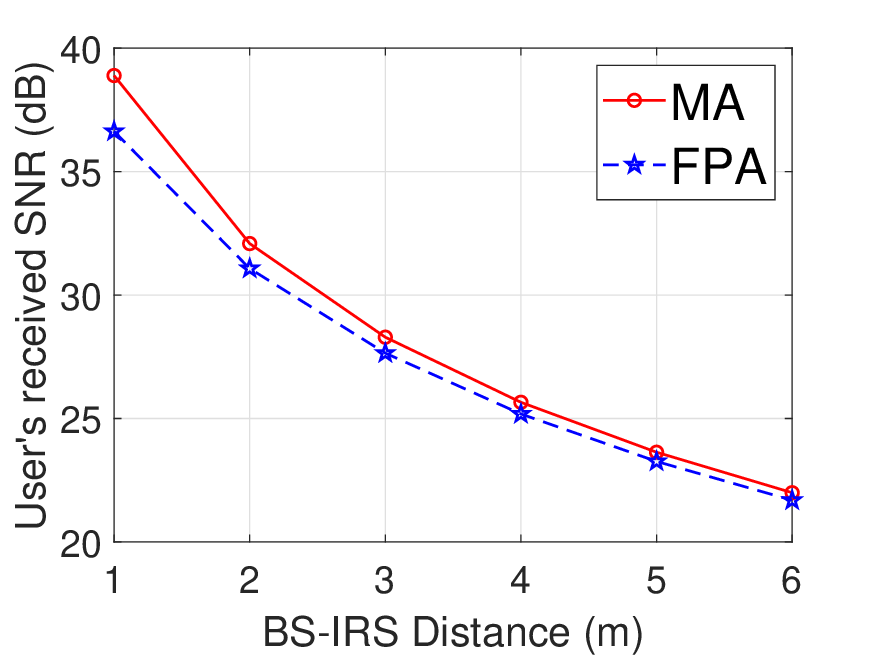}
		}
		\captionsetup{font=footnotesize}
		\caption{Received SNR at the user with (a) a single MA; (b) $N=4$ MAs under the LoS BS-IRS channel.}
		\vspace{-15pt}
	\end{figure}
	
	To verify the above analyses, we plot in Fig. 2(a) the received SNR at the user versus the BS-IRS distance under the near-field LoS BS-IRS channel. The BS is equipped with a single-MA. The operating frequency is $f=5$ GHz. The total number of the IRS reflecting elements is $M=25^2$. The transmit SNR is $P/\sigma^2=110$ dB. In the FPA benchmark scheme, the antenna is fixed at \eqref{eqn_NearestPos}. It is observed that the SNR performance by a single MA and a single FPA are identical for all BS-IRS distances considered, which validates our analyses.
	
	\begin{remark}
		Different from the above case with an IRS, if we consider a near-field LoS channel from the single-MA BS to the user, the BS-user channel is given by
		\begin{equation}\label{eqn_Ch_BU}
			h_{\text{BU}}(\bs{t})=\frac{\lambda}{4\pi||\bs{t}-\bs{q}_U||}e^{j\frac{2\pi}{\lambda}||\bs{t}-\bs{q}_U||},
		\end{equation}
		where $\bs{q}_U=[x_U,y_U,z_U]^T\ib{R}^{3\times1}$ denotes the coordinate of the user. It can be verified that the optimal antenna position that maximizes \eqref{eqn_Ch_BU} is given by
		\begin{equation}
			\bs{t}^{\star} = \arg\underset{\bs{t}\ic{C}_t}{\min}\,\,||\bs{t}-\bs{q}_U||.
		\end{equation}
		Considering that $\ca{C}_t$ is parallel to the $x$-axis, we have
		\begin{equation}
			\boldsymbol{t}^{\star}=\begin{cases}
				\left[ -\frac{A}{2}+x_B,y_B,z_B \right] ^T,\mathrm{if}\,\, x_U\le -\frac{A}{2},\\
				\left[ x_U,y_B,z_B \right] ^T,\quad\qquad\mathrm{if}\,\, -\frac{A}{2}<x_U\le \frac{A}{2},\\
				\left[ \frac{A}{2}+x_B,y_B,z_B \right] ^T,\,\,\,\,\mathrm{if}\,\, x_U>-\frac{A}{2},\\
			\end{cases}
		\end{equation}
		which depends on the user's coordinate $\bs{q}_U$ (or $x_U$) unlike the IRS-aided case.
	\end{remark}
	
	\subsection{Fluctuation of Channel Power Gain}
	In this subsection, we analyze the fluctuation of \eqref{eqn_RecPow_Approx} within $\ca{C}_t$. Notably, a more significant variation of the channel power gain generally results in a more pronounced spatial diversity within $\ca{C}_t$ (and hence the performance gain of MAs over FPAs). To characterize the fluctuation, we derive the difference of \eqref{eqn_RecPow_Approx} with respect to (w.r.t.) two arbitrary locations in $\ca{C}_t$, denoted as $\bs{t}_1$ and $\bs{t}_2$ ($\bs{t}_1\ne\bs{t}_2$). Without loss of generality, we assume that $\bs{t}_1$ is closer to the origin than $\bs{t}_2$, i.e., $R(\bs{t}_1) < R(\bs{t}_2)$. Then, the difference between the maximum channel power gains at these two positions is given by
	\begin{equation}
		\small
		\begin{aligned}\label{eqn_Pu_Diff}
			G_u(\bs{t}_1)-G_u(\bs{t}_2)&=H^2(M_y,M_z,\bs{t}_1)-H^2(M_y,M_z,\bs{t}_2)\\
			&=\left[H(M_y,M_z,\bs{t}_1)+H(M_y,M_z,\bs{t}_2)\right]\\
			&\quad\times\left[H(M_y,M_z,\bs{t}_1)-H(M_y,M_z,\bs{t}_2)\right],
		\end{aligned}
	\end{equation}
	where the constant scalar $\left(\frac{\lambda}{4\pi }\right)^2$ is omitted for simplicity.
	
	Based on \eqref{eqn_Pu_Diff}, we analyze the effects of different key system parameters on it. First, it is easy to see that $H(M_y,M_z,\bs{t}_1)\pm H(M_y,M_z,\bs{t}_2)$ monotonically increases with $M_y$ and $M_z$, provided that $R(\bs{t}_1) < R(\bs{t}_2)$. Thus, $G_u(\bs{t}_1)-G_u(\bs{t}_2)$ also monotonically increases with $M_y$ and $M_z$, for any given $\bs{t}_1$ and $\bs{t}_2$. This indicates that by increasing the size of the IRS, the maximum channel power gains within $\ca{C}_t$ experience more significant fluctuation.
	
	On the other hand, we analyze the effects of the BS-IRS distance on \eqref{eqn_Pu_Diff}. For simplicity, we assume that the MA array is perpendicular to the IRS. Then, $R(\bs{t}_i)$ can be expressed as $R(\bs{t}_i)=\sqrt{d_{\text{BI}}^2+d^2(\bs{t}_i)}$, $i=1,2$, where $d(\bs{t}_i)=||\bs{q}_B-\bs{t}_i||$ denotes the distance between the $i$-th location and the center of the MA array, and $d_{\text{BI}}$ denotes the BS-IRS distance. As $R(\bs{t}_1)<R(\bs{t}_2)$, $H(M_y,M_z,\bs{t}_1)\pm H(M_y,M_z,\bs{t}_2)$ monotonically decreases with $d_{\text{BI}}$. Thus, $G_u(\bs{t}_1)-G_u(\bs{t}_2)$ also decreases with $d_{\text{BI}}$. This indicates that a more/less significant fluctuation of \eqref{eqn_RecPow_Approx} will be resulted if the BS-IRS distance reduces/increases.
	
	Based on the above, although the antenna position yielding the maximum channel power gain is fixed as \eqref{eqn_NearestPos}, the channel power gain within $\ca{C}_t$ may vary. As such, it can be inferred that in the case of multiple antennas, the performance gain of MAs over FPAs may still exist, unlike the single-MA case. To verify this claim, we plot in Fig. 2(b) the received SNR at the user versus the BS-IRS distance with $N=4$ MAs and other simulation parameters the same as Fig. 2(a). It is observed that different from the single-MA case, employing multiple MAs can still yield a performance gain over FPAs, if the BS-IRS distance is small. Moreover, the performance gain is observed to decrease with the BS-IRS distance, which is consistent with our previous analyses.
	
	\subsection{Far-Field BS-IRS Channel}
	All of the above analytical results are derived under the general near-field BS-IRS channel model. Next, we consider a special case with a far-field BS-IRS channel, under which the distance between the MA and the $(m_y,m_z)$-th IRS reflecting element can be treated as identical, i.e.,
	\begin{equation}\label{eqn_Dist_Approx}
		\begin{aligned}
			D(\bs{t},m_y,m_z)&\approx\sqrt{R^2(\bs{t})+(m_yd)^2+(m_zd)^2}\\
			&\approx d_{\text{BI}},\,\,\forall m_y,m_z.
		\end{aligned}
	\end{equation}
	By substituting \eqref{eqn_Dist_Approx} into \eqref{eqn_Pu_Diff}, it can be seen that $G_u(\bs{t}_1)-G_u(\bs{t}_2)=0$, $\forall \bs{t}_1, \bs{t}_2$. This implies that in the far-field scenario, the channel power gain within $\ca{C}_t$ is uniform. As a result, even employing multiple MAs may not achieve any performance gain over FPAs, as can be rigorously shown below.
	
	Given a multi-MA BS, the BS-IRS channel can be expressed as $\bs{H}_{\text{BI}}(\bs{T})=\beta_{\text{BI}}\bs{u}\bs{v}^H(\bs{T})$, where $\beta_{\text{BI}}$ encompasses the large-scale path loss, $\bs{u}\ib{C}^{M\times1}$ and $\bs{v}(\bs{T})\ib{C}^{N\times1}$ denote the receive and transmit array responses at the IRS and BS, respectively. Accordingly, the channel power gain can be expressed as $G_u(\bs{T},\bs{\Psi})=\left|\beta_{\text{BI}}|^2|\bs{h}_{\text{IU}}\bs{\Psi}\bs{u}\right|^2\left|\bs{v}^H(\bs{T})\bs{w}\right|^2$. It can be seen that for any given $\bs{T}$, to maximize $G_u(\bs{T},\bs{\Psi})$, the optimal BS active beamforming should be set as 
	\begin{equation}
		\bs{w}(\bs{T})=\frac{\bs{v}(\bs{T})}{||\bs{v}(\bs{T})||},
	\end{equation}
	leading to $\left|\bs{v}^H(\bs{T})\bs{w}\right|^2=N$, regardless of $\bs{T}$. Hence, employing multiple MAs cannot yield any performance gain over FPAs in the far-field scenario.
	
	It is noteworthy that we have previously shown in Section III-B that multiple MAs can achieve a more substantial performance gain over FPAs as the BS-IRS distance decreases and/or the IRS size increases. This in fact renders the BS-IRS channel condition closer to the near field. Consequently, it can be concluded that the performance gains of multiple MAs over multiple FPAs may become increasingly pronounced as the BS-IRS channel becomes more dominated by near-field propagation.

	\section{Proposed Solution to (P1)}
	In this section, we focus on the general BS-IRS channel and solve (P1). Note that for any given APV $\bs{T}$ and IRS passive beamforming $\bs{\Psi}$, the optimal BS transmit beamforming is given by the maximum transmission ratio (MRT), i.e.,
	\begin{equation}\label{eqn_MRT}
		\bs{w}(\bs{T},\bs{\Psi})=\frac{\left(\bs{h}_{\text{IU}}^H\bs{\Psi}\bs{H}_{\text{BI}}(\bs{T})\right)^H}{||\bs{h}_{\text{IU}}^H\bs{\Psi}\bs{H}_{\text{BI}}(\bs{T})||}.
	\end{equation}
	By substituting \eqref{eqn_MRT} into (P1), it can be expressed as
	\begin{subequations}
		\begin{align}
			{\text{(P1)}}\quad &\underset{\bs{\Psi},\bs{T}}{\max}\quad \gamma(\bs{T},\bs{\Psi})=||\bs{h}_{\text{IU}}^H\bs{\Psi}\bs{H}_{\text{BI}}(\bs{T})||^2 \nonumber
			\\
			\mathrm{s.t.}\quad & \text{\eqref{eqn_IRS_Phase_Cons}},\,\, \text{\eqref{eqn_MA_Region_Cons}},\,\, \text{\eqref{eqn_MA_Coordinate_Cons}}, \nonumber
		\end{align}
	\end{subequations}
	where the constant $P/\sigma^2$ is omitted. However, (P1) is still a non-convex optimization problem. Next, we propose an AO algorithm to decompose (P1) into two subproblems and solve them accordingly.
	
	\subsection{Optimizing $\bs{\Psi}$ for Given $\bs{T}$}
	First, we optimize the IRS reflection matrix $\bs{\Psi}$ for any given APV $\bs{T}$. Let $\bs{G}_1\triangleq\text{diag}(\bs{h}_\text{IU}^H)\bs{H}_{\text{BI}}(\bs{T})\ib{C}^{M\times N}$ and $\bs{g}_{1,m}\ib{C}^{1\times N}$ denote the $m$-th ($m\ic{M}$) row of $\bs{G}_1$, which are fixed with a given APV $\bs{T}$. Then, (P1) can be simplified as
	\begin{subequations}\label{eqn_OptPrblm_P2}
		\begin{align}
			{\text{(P2)}}\quad &\underset{\bs{\varphi}}{\max}\quad \gamma(\bs{T},\bs{\Psi})=\left|\left|\sum_{m=1}^{M}{\bs{g}_{1,m}e^{j\varphi_m}}\right|\right|^2 \nonumber
			\\
			\mathrm{s.t.}\quad & \varphi_m\in[0,2\pi],\,\,\forall m \ic{M},
		\end{align}
	\end{subequations}
	which, however, is still a non-convex problem. To tackle this challenge, we utilize an element-wise block coordinate descent (BCD) method to optimize the IRS reflecting coefficient sequentially. Specifically, it can be easily shown that for any given $\varphi_i$, $i \in {\cal M}$, $i \ne m$, the optimal $\varphi_m$ for (P2) is given by
	\begin{equation}\label{eqn_OptPhase}
		\varphi_m^{\star}=\arg\left(\bs{\alpha}_m\bs{g}_{1,m}^H\right),
	\end{equation}
	where $\bs{\alpha}_m=\sum_{i\ne m}^{M}{\bs{g}_{1,i}e^{-j\varphi_i}}$. As such, in the $m$-th BCD iteration, we can fix $\varphi_i$, $i \ne m$, $i \in {\cal M}$ and optimize $\varphi_m$ as \eqref{eqn_OptPhase}. Then, the BCD iteration proceeds until the phase shifts of all $M$ IRS reflecting elements have been updated.
	
	\subsection{Optimizing $\bs{T}$ for Given $\bs{\Psi}$}
	Next, we optimize the APV $\bs{T}$ with a given IRS reflection matrix $\bs{\Psi}$. Let $\bs{h}_{\text{BI}}(\bs{t}_n)\ib{C}^{M\times1}$ denote the $n$-th column of $\bs{H}_{\text{BI}}(\bs{T})$. Then, the objective function of (P1), i.e., \eqref{eqn_SNR}, can be recast as
	\begin{equation}
		\begin{aligned}
			\gamma(\bs{T},\bs{\Psi})=\sum_{n=1}^{N}{\left|\bs{g}_2^H\bs{h}_{\text{BI}}(\bs{t}_n)\right|^2},
		\end{aligned}
	\end{equation}
	where $\bs{g}_2^H=\bs{h}_{\text{IU}}^H\bs{\Psi}$. As such, (P1) can be simplified as
	\begin{subequations}\label{eqn_OptPrblm_P3}
		\begin{align}
			{\text{(P3)}}\quad &\underset{\bs{T}}{\max}\,\, \sum_{n=1}^{N}{\left|\bs{g}_2^H\bs{h}_{\text{BI}}(\bs{t}_n)\right|^2} \quad\mathrm{s.t.}\,\,\,\, \text{\eqref{eqn_MA_Region_Cons}},\,\,\text{\eqref{eqn_MA_Coordinate_Cons}},\nonumber
		\end{align}
	\end{subequations}
	which can be solved \emph{optimally} by applying the graph-based approach proposed in our previous work \cite{W_Mei_Graph} with modifying the weight assignment therein.
	
	Specifically, we uniformly sample the transmit array into $L_{\text{samp}}$ discrete sampling points with an equal spacing $\delta_s=A/L_{\text{samp}}$ between any two adjacent sampling points. Let $s_l$, $s_l\ib{R}^{3\times1}$ denote the coordinate of the $l$-th sampling point, $l\ic{L}\triangleq\{1,2,\cdots,L_\text{samp}\}$, and $a_n$ denote the index of the selected sampling point for the $n$-th MA, $a_n\ic{L}$. Then, we can transform (P3) into the following discrete point selection problem:
	\begin{subequations}\label{eqn_OptPrblm_P4}
		\begin{align}
			{\text{(P4)}}\quad &\underset{\left\{a_n\right\}}{\max}\,\, \sum_{n=1}^{N}{\left|g_{a_n}\right|^2}\nonumber\\
			&\,\,\mathrm{s.t.}\,\,\,\, a_n\ic{L},\\
			&\,\,\quad\quad|a_i-a_j|>a_{\min}, \forall i,j\ic{N}, i\ne j,
		\end{align}
	\end{subequations}
	where $g_{a_n}=\bs{g}_2^H\bs{h}_{\text{BI}}(s_{a_n})$, $\forall n\ic{N}$, and $a_{\min}=D_{\min}/\delta_s\gg1$. Next, we can construct a directed weighted graph to equivalently transform (P4) into a fixed-hop shortest path problem that can be optimally solved in polynomial time using the dynamic programming. The details are omitted for brevity, for which interested readers can refer to \cite{W_Mei_Graph}. It is worth noting that the graph-based algorithm is general, as it eliminates reliance on specific channel models for MAs.
	
	Based on above, we can alternately solve (P2) and (P4) by applying the element-wise BCD method and the graph-based approach. As this process always yields a non-decreasing objective value of (P1), the convergence of AO is guaranteed.
	
	\section{Numerical Results}
	In this section, we provide numerical results to evaluate the efficacy of our proposed AO algorithm. Unless other specified, the simulation parameters are set as follows. The operating frequency is $f=5$ GHz. The number of the transmit MAs is $N=4$. The total number of the IRS reflecting elements along the $y$- and $z$-axes is set identical as $M_y=M_y=15$, leading to $M=M_y\times M_z = 225$. The spacing between any two adjacent IRS reflecting elements is $d=\frac{\lambda}{2}$, and the minimum spacing between any two MAs at the BS is $D_{\min}=\frac{\lambda}{2}$. The length of the transmit region $\ca{C}_t$ is $A=0.6$ m. The spacing of the sampling points is $\delta_s=0.01$ m. The transmit SNR is $P/\sigma^2=110$ dB, and the coordinate of the center of the transmit region $\ca{C}_t$ is $\bs{q}_B=[5,5,0]^T$. We consider the Rician-fading IRS-user channel with the Rician factor set to $K=3$ dB. The distance from the IRS to the user is  $d_{\text{IU}}=30$ m, and the path loss exponent for the IRS-user channel is $\kappa=2.8$. Moreover, we consider the near-field multi-path BS-IRS channel. Let $L$ denote the number of dominant paths over the BS-IRS link, and $\bs{s}_{\text{BI},l}\ib{R}^{3\times1}$ denote the coordinate of the $l$-th scatterer, $l=1,2,\cdots,L$. Then, the BS-IRS channel can be expressed as  \cite{Y_Liu_NF_Review}
	\begin{equation}\label{eqn_Channel_BI_NF}
		\bs{H}_{\text{BI}}(\bs{T})=\underset{\text{LoS component}}{\underbrace{\bs{H}_{\text{BI}}^{\text{LoS}}(\bs{T})}}+\underset{\text{NLoS components}}{\underbrace{\sum_{l=1}^{L}{\beta_{\text{BI},l}\bs{a}^T(\bs{s}_{\text{BI},l})\bs{b}(\bs{T},\bs{s}_{\text{BI},l}})}},
	\end{equation}
	where the LoS component $\bs{H}_{\text{BI}}^{\text{LoS}}(\bs{T})$ is written as
	\begin{equation}
		\small
		\bs{H}_{\text{BI}}^{\text{LoS}}(\bs{T})=\left[\frac{\lambda}{4\pi D(\bs{t}_n,m_y,m_z)}e^{j\frac{2\pi}{\lambda}D(\bs{t}_n,m_y,m_z)}\right]_{m_y\ic{M}_y,m_z\ic{M}_z}^{n\ic{N}},
	\end{equation}
	with $D(\bs{t}_n,m_y,m_z)=||\bs{t}_n-\bs{e}_{m_y,m_z}||$ denoting the distance between the $n$-th MA and the $(m_y,m_z)$-th IRS reflecting element, $\forall n,m_y,m_z$. In addition, the vectors $\bs{a}(\bs{s}_{\text{BI},l})\ib{C}^{1\times M}$ and $\bs{b}(\bs{T},\bs{s}_{\text{BI},l})\ib{C}^{1\times N}$ denote the receive and transmit near-field array responses at the IRS and the BS, respectively, which are given by
	\begin{subequations}
		\begin{align}
			\bs{a}(\bs{s})&=\left[e^{j\frac{2\pi}{\lambda}||\bs{s}-\bs{e}_{m_y,m_z}||}\right]_{m_y\ic{M}_y,m_z\ic{M}_z},\\
			\bs{b}(\bs{T},\bs{s})&=\left[e^{j\frac{2\pi}{\lambda}||\bs{s}-\bs{t}_n||}\right]_{n\ic{N}}.\label{eqn_NF_Vec}
		\end{align}
	\end{subequations}
	
	Furthermore, we consider the following benchmark schemes for performance comparison:
	\begin{enumerate}
		\item  \textbf{FPAs with antenna selection (FPAs w/ AS)}: In this benchmark, $A/D_{\min}$ FPAs are deployed within $\ca{C}_t$ and separated by the minimum distance $D_{\min}$. Among them, $N$ antennas are selected for transmission. The associated optimization problem can be solved by applying a similar AO algorithm as in Section IV by setting $\delta_s=D_{\min}$.
		
		\item \textbf{FPAs without antenna selection (FPAs w/o AS)}: The $N$ MAs are deployed symmetrically to $\bs{q}_B$ and separated by the minimum distance $D_{\min}$. The IRS reflection matrix is optimized as in Section IV-A.
	\end{enumerate}
	
	\begin{figure}[!t]
		\centering
		\captionsetup{justification=raggedright,singlelinecheck=false}
		\includegraphics[width=0.78\linewidth]{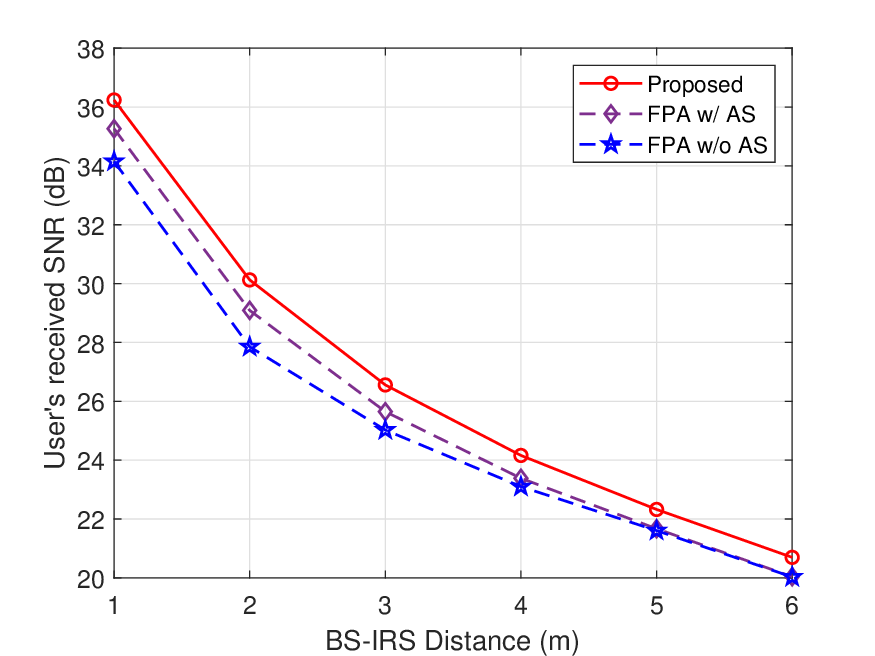}
		\captionsetup{font=footnotesize}
		\caption{Received SNR at the user versus the BS-IRS distance.}
		\vspace{-15pt}
	\end{figure}
	
	\begin{figure}[!t]
		\centering
		\captionsetup{justification=raggedright,singlelinecheck=false}
		\includegraphics[width=0.78\linewidth]{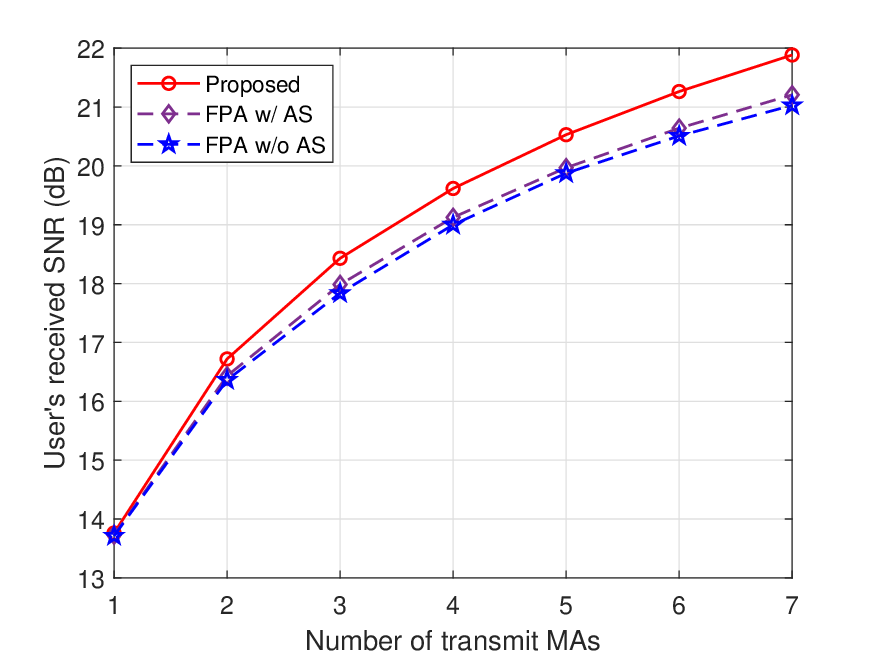}
		\captionsetup{font=footnotesize}
		\caption{Received SNR at the user versus the number of transmit MAs.}
		\vspace{-15pt}
	\end{figure}
	
	\begin{figure}[!t]
		\centering
		\captionsetup{justification=raggedright,singlelinecheck=false}
		\includegraphics[width=0.78\linewidth]{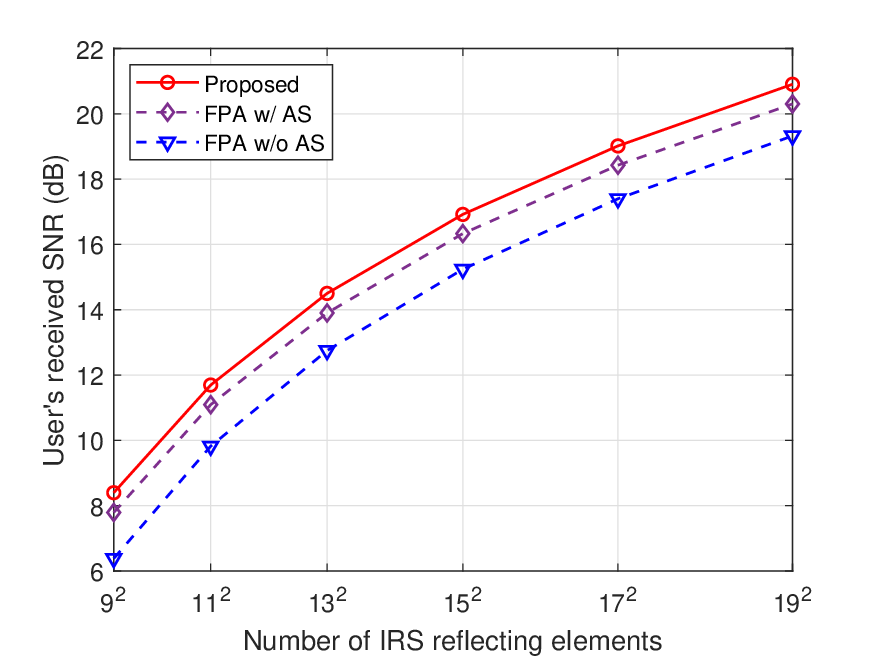}
		\captionsetup{font=footnotesize}
		\caption{Received SNR at the user versus the number of IRS reflecting elements.}
		\vspace{-15pt}
	\end{figure}
	
	First, we plot in Fig. 3 the received SNR at the user versus the BS-IRS distance, with $N=4$ and $L=8$. It is observed that our proposed algorithm achieves higher received SNRs than the other two benchmarks, thanks to the more flexible antenna position optimization. In addition, it can be seen that the performance gain of our proposed algorithm over the FPA w/o AS decreases with the BS-IRS distance, which shows a similar trend to the case with a LoS BS-IRS channel in Fig. 2(b). Nonetheless, the performance gain is observed to remain (at around 1 dB) as the BS-IRS distance increases, rather than becoming negligible as shown in Fig. 2(b). This implies that in the case of the multi-path BS-IRS channel, employing MAs can lead to a larger performance gain over FPAs compared to the case of LoS BS-IRS channel, whether in the far field or near field. However, the performance gain is limited to around 1-2 dB over the whole range of BS-IRS distances considered. This may be due to IRS passive beamforming sufficiently reconfiguring the end-to-end channel.
	
	Next, we plot in Fig. 4 the received SNR at the user versus the number of the transmit MAs (i.e., $N$) with $L=8$. It is observed that the received SNRs by all schemes increase with the number of MAs, thanks to the enlarged beamforming gain. Moreover, as $N$ increases, employing MAs is observed to yield a more significant gain over the two benchmarks, implying that more spatial diversity gain can be exploited for a larger $N$ in the case of the multi-path BS-IRS channels.
	
	Lastly, we plot in Fig. 5 the received SNR versus the number of IRS reflecting elements, i.e., $M$, with $N=4$ and $L=8$. It is observed that the received SNRs by all schemes increase with $M$, as more reflecting elements can provide more significant passive beamforming gain. However, the performance gain of the proposed AO algorithm over the FPA w/o AS slightly decreases with $M$, e.g., from 2.01 dB at $M=9^2$ to 1.58 dB at $M=19^2$. This observation is also similar to the trend for the BS-IRS LoS channel discussed in Section III-B.
	
	\section{Conclusions}
	In this paper, we studied a joint active/passive beamforming and MA position optimization problem for an MA-enhanced IRS-assisted wireless communication system and proposed an AO algorithm to solve it. Our performance analyses shows that under the LoS BS-IRS channel and optimal IRS passive beamforming, the performance gain of MAs over FPAs vanishes in the case of a single MA or far-field BS-IRS propagation, but increases as the BS-IRS channel approaches the near-field. Numerical results also revealed a similar trend for multi-path BS-IRS channels and that MAs achieve more significant gains over FPAs compared to LoS BS-IRS channels.

\end{document}